\let\orilabel\label
\documentclass[aps,pra,preprint,a4paper,showpacs,showkeys,superscriptaddress,nofootinbib]{revtex4-1}
\let\label\orilabel

\usepackage{latexsym}
\usepackage{amsmath,amssymb,mathrsfs}
\usepackage{mathtools}
\usepackage{graphicx}
\usepackage{subfigure}
\usepackage{xcolor}
\usepackage{physics}
\usepackage{cancel}
\usepackage{easyReview}
\usepackage{tikz}
\usepackage{multirow,tabularx}

\newcolumntype{C}[1]{>{\centering\arraybackslash}p{#1}}
\usepackage{hyperref}     
\hypersetup{colorlinks,%
  citecolor=blue,%
  linkcolor=cyan,%
}
\usepackage[titletoc]{appendix}
\usepackage{enumerate}
\graphicspath{{./Figures/}}
\usetikzlibrary{decorations.pathmorphing,decorations.pathreplacing,decorations.shapes}
\begin{document}


\title{Gravitational constant as a conserved charge in black hole thermodynamics}

\author{Wontae Kim}%
\email[]{wtkim@sogang.ac.kr}%
\affiliation{Department of Physics, Sogang University, Seoul, 04107,
	Republic of Korea}%
\affiliation{Center for Quantum Spacetime, Sogang University, Seoul 04107, Republic of Korea}%

\author{Mungon Nam}%
\email[]{clrchr0909@sogang.ac.kr}%
\affiliation{Center for Quantum Spacetime, Sogang University, Seoul 04107, Republic of Korea}%
\date{\today}

\begin{abstract}
Recent work has demonstrated that coupling constants in a given action can be promoted to the role of conserved charges.  This is achieved by introducing pairs of field variables constructed from combinations of scalar and gauge fields. This framework naturally suggests that the gravitational constant itself can be interpreted as a conserved charge, arising from an associated gauge symmetry. In a modified four-dimensional Einstein-Hilbert action, we explicitly show that the gravitational constant, in addition to the mass and the cosmological constant, emerges as a conserved charge. Our derivation, which employs the quasi-local off-shell ADT formalism, yields a result that is fully consistent with the extended thermodynamic first law and the Smarr formula.
\end{abstract}
%



\maketitle


\raggedbottom

\section{Introduction}
\label{sec:introduction}
In classical general relativity, stationary black holes are uniquely characterized by three conserved quantities: mass, electric charge, and angular  momentum~\cite{Israel:1967wq,Israel:1967za,Carter:1971zc,Robinson:1975bv}. These quantities, which arise as integration constants in the equations of motion, correspond to the global charges of spacetime. By treating these charges as thermodynamic variables, along with entropy and temperature, one can formulate the four laws of black hole thermodynamics~\cite{Bardeen:1973gs,Bekenstein:1973ur,Hawking:1975vcx}.
In fact, there have been numerous proposals including certain parameters in the gravitational action as thermodynamic variables. For example, in anti-de Sitter (AdS) spacetime, the cosmological constant can be interpreted as a thermodynamic pressure with its conjugate variable being a thermodynamic volume, leading to an extended thermodynamic first law~\cite{Henneaux:1984ji,Henneaux:1989zc,Teitelboim:1985dp,Bhattacharya:2017hfj}. In this extended framework, often referred to as ``black hole chemistry,'' the black hole mass is identified with the enthalpy. The resulting phase structures and thermodynamic properties have been extensively studied in Refs.~\cite{Kastor:2009wy,Cvetic:2010jb,Dolan:2010ha,Dolan:2011xt,Frassino:2015oca,Kubiznak:2016qmn,Ortin:2021ade,Mann:2025xrb}.

From the holographic perspective of the AdS/CFT correspondence~\cite{Maldacena:1997re,Witten:1998qj}, the cosmological  and the gravitational constants are related to the number of degrees of freedom in the dual conformal field theory. This observation has motivated a holographic reinterpretation of black hole chemistry, wherein the central charge is treated as a thermodynamic variable with an associated chemical potential~\cite{Karch:2015rpa,Visser:2021eqk,Cong:2021jgb,Ahmed:2023dnh,Ahmed:2023snm,Cong:2021fnf,AlBalushi:2020rqe,Mancilla:2024spp}. Meanwhile, the thermodynamics of AdS black holes has been investigated by varying the gravitational constant while holding the cosmological constant fixed~\cite{Zeyuan:2021uol,Gao:2021xtt,Wang:2021cmz,Volovik:2021iim,Ali:2023ppg,Volovik:2025xmq}.
Thus, a fundamental question naturally arises as to how the cosmological and the gravitational constants, which are fixed parameters in the action, can be consistently treated as thermodynamic variables. The answer lies in establishing a principle that legitimises their variation in the thermodynamic first law. It is indeed possible to promote the cosmological constant to an integration constant of the equations of motion, thereby a conserved charge, by introducing an auxiliary field that implements a gauge symmetry~\cite{Aurilia:1980xj,Duff:1980qv,Hawking:1984hk,Bousso:2000xa,Chernyavsky:2017xwm,Hajian:2021hje}.
More recently, it was shown that any arbitrary coupling in an action can be promoted to a conserved charge associated with a gauge symmetry by introducing a scalar field paired with a
$(D-1)$-form gauge field~\cite{Meessen:2022hcg,Zatti:2023oiq,Ballesteros:2023muf,Hajian:2023bhq}.

In this paper, adopting the approach in Refs.~\cite{Meessen:2022hcg,Zatti:2023oiq,Ballesteros:2023muf,Hajian:2023bhq}, we explicitly demonstrate that the gravitational constant can be interpreted as a conserved charge in the modified Einstein-Hilbert action. Based on the Abbott-Deser-Tekin (ADT) formalism~\cite{Abbott:1981ff,Deser:2002rt,Deser:2002jk,Adami:2017phg}, we employ the off-shell framework which provides a quasi-local construction of conserved charges by expressing the ADT potential in terms of the off-shell linearized Noether potential~\cite{Kim:2013zha}. Using the derived quasi-local charges, we obtain the thermodynamic first law and the Smarr formula~\cite{Smarr:1972kt}.

The organization of this paper is as follows. In Sec.~\ref{sec:hh}, we introduce a modified Einstein-Hilbert action coupled to two scalar-gauge pairs, which allows the promotion of both the gravitational constant $G$ and the cosmological constant $\Lambda$ to integration constants. We then apply the off-shell ADT formalism to obtain the quasi-local conserved charges associated with the underlying gauge symmetries. In Sec.~\ref{sec:extended}, we derive the extended thermodynamic first law and the Smarr formula relating these charges. Finally, conclusion and discussion will be provided in Sec.~\ref{sec:conclusion}.

\section{Quasi-local conserved charges}
\label{sec:hh}
We start with the four-dimensional action described by
\begin{equation}
	\label{eq:action}
S=\int \dd[4]x\sqrt{-g}\mathcal{L} = \int \dd[4]x\sqrt{-g}\alpha \left[ R  + \beta\left( 1-\nabla_{\mu}B^\mu \right) - \nabla_{\mu}A^\mu \right],
\end{equation}
where $\alpha$ and $\beta$ are scalar fields, and $A^\mu$ and $B^\mu$ are gauge fields~\cite{Hajian:2023bhq}.
Varying the action \eqref{eq:action} with respect to $g_{\mu\nu}$, $\alpha$, $\beta$, $A^\mu$, and $B^\mu$ yields
\begin{equation}
	\label{eq:action var}
	\delta S = \int \dd[4] x \left[ \sqrt{-g}\left( -\mathcal{E}^{\mu\nu} \delta g_{\mu\nu} + \mathcal{E}_\alpha \delta \alpha + \mathcal{E}_\beta\delta\beta + \mathcal{E}^A_\mu \delta A^\mu + \mathcal{E}^B_\mu\delta B^\mu\right)  + \partial_{\mu}\Theta^\mu(\delta g, \delta A, \delta B)\right],
\end{equation}
where
\begin{gather}
	\mathcal{E}^{\mu\nu} = \alpha R^{\mu\nu}-\nabla^{\mu}\nabla^{\nu}\alpha - \frac{1}{2}g^{\mu\nu}\left[ \alpha\left(  R + \beta \right) + \left( A^{\lambda} + B^{\lambda} \right)\nabla_{\lambda}\alpha + B^{\lambda}\nabla_{\lambda}\beta - 2\square \alpha \right],\label{eq:eom g}\\
	\mathcal{E}_\alpha = R + \beta\left( 1-\nabla_{\mu}B^\mu \right) - \nabla_{\mu}A^{\mu} ,\quad  \mathcal{E}_\beta = \alpha\left( 1- \nabla_{\mu}B^{\mu} \right) ,\quad \mathcal{E}^A_\mu = \nabla_{\mu}\alpha,\quad \mathcal{E}^B_\mu = \nabla_{\mu}\left( \alpha\beta \right),\label{eq:eom aux}
\end{gather}
and the surface term is
\begin{equation}
	\label{eq:surface term}
	\Theta^\mu(\delta g, \delta A, \delta B) = 2\sqrt{-g}g^{\mu[\kappa}g^{\lambda]\nu}\left[ \alpha\nabla_{\lambda}\delta g_{\nu\kappa}-\nabla_{\lambda}\alpha \delta g_{\nu\kappa} \right] -  \alpha \left[ \delta\left( \sqrt{-g}A^\mu \right) - \beta \delta\left( \sqrt{-g}B^\mu \right) \right].
\end{equation}
Solving the field equations~\eqref{eq:eom g} and \eqref{eq:eom aux}, one can find a spherically symmetric static solution in terms of ingoing Eddington–Finkelstein coordinates $(v,r,\theta,\phi)$,
\begin{equation}
	\label{eq:metric solution}
	\dd s^2 = - \left( 1-\frac{\gamma_0}{r}+\frac{1}{6 }\beta_0 r^2 \right) \dd v^2 + 2\dd v\dd r + r^2 \dd \Omega^2,\quad
\end{equation}
with the other solutions as
\begin{equation}
	\label{eq:field solutions}
	\alpha = \alpha_0,\quad \beta = \beta_0,\quad A^r = -\frac{2}{3} \beta_0  r+ \frac{C}{r^2},\quad B^r = \frac{1}{3}r + \frac{D}{r^2},\\
\end{equation}
where $\dd\Omega^2$ is the line element on the unit two-sphere, and $\alpha_0$, $\beta_0$, $\gamma_0$, $C$, and $D$ are integration constants.

In Eq.~\eqref{eq:action}, under the infinitesimal diffeomorphism generated by $\zeta^\mu$, the metric transforms as $\mathcal{L}_{\zeta} g_{\mu\nu} = \nabla_{\mu}\zeta_{\nu} + \nabla_{\nu}\zeta_{\mu}$. The action is also invariant under the local gauge transformations:
\begin{equation}
	\label{}
	A^\mu \longrightarrow A^\mu + \nabla_{\nu}\lambda^{\mu\nu},\quad B^{\mu}\longrightarrow B^\mu + \nabla_{\nu}\chi^{\mu\nu},
\end{equation}
where $\lambda^{\mu\nu}$ and $\chi^{\mu\nu}$ are arbitrary antisymmetric local parameters.
In connection with these symmetries, we identify the corresponding off-shell Noether current by considering the combined variation consisting of a diffeomorphism and the two independent gauge transformations defined as
\begin{gather}
	\delta g_{\mu\nu} = \nabla_{\mu}\zeta_{\nu} + \nabla_{\nu}\zeta_{\mu},\quad \delta\alpha = \zeta^\mu \nabla_{\mu}\alpha,\quad \delta\beta = \zeta^\mu \nabla_{\mu}\beta,\label{eq:field var}\\
	\delta A^\mu = \zeta^\nu \nabla_{\nu}A^\mu - A^\nu \nabla_{\nu}\zeta^\mu + \nabla_{\nu}\lambda^{\mu\nu},\quad \delta B^\mu = \zeta^\nu \nabla_{\nu}B^\mu - B^\nu \nabla_{\nu}\zeta^\mu + \nabla_{\nu}\chi^{\mu\nu}.\label{eq:vector var}
\end{gather}
Inserting Eqs.~\eqref{eq:field var} and \eqref{eq:vector var} into Eq.~\eqref{eq:action var} with the relation $\mathcal{L}_{\zeta}\left( \sqrt{-g}\mathcal{L} \right) = \partial_{\mu}\left(  \sqrt{-g}\zeta^\mu\mathcal{L} \right)$,
we can find
\begin{equation}
	\label{total}
	\int \dd[4]x \partial_{\mu}\left[ \sqrt{-g}\left( -2\mathcal{E}^{\mu\nu}\zeta_{\nu} - \mathcal{E}_{\nu}^{A}\left( A^{\mu}\zeta^\nu  + \lambda^{\mu\nu} \right) - \mathcal{E}^{B}_{\nu}\left( B^{\mu}\zeta^{\nu} + \chi^{\mu\nu}\right) - \zeta^{\mu}\mathcal{L} \right) + \Theta^{\mu}(\zeta,\lambda,\chi) \right]=0,
\end{equation}
where we used the relations $\lambda^{\mu\nu}\nabla_{\mu}\nabla_{\nu} \alpha = \chi^{\mu\nu}\nabla_{\mu}\nabla_{\nu} \beta = 0$ and the off-shell Noether (Bianchi-like) identity written as
\begin{equation}
	\label{eq:Bianchi}
	2\nabla^{\nu}\mathcal{E}_{\mu\nu}+ \mathcal{E}_{\alpha}\nabla_{\mu}\alpha + \mathcal{E}_{\beta}\nabla_{\mu}\beta +\mathcal{E}_{\nu}^{A}\nabla_{\mu}A^{\nu} + \nabla_{\nu}\left(\mathcal{E}_{\mu}^{A} A^{\nu} \right) + \mathcal{E}_{\nu}^{B}\nabla_{\mu}B^{\nu} + \nabla_{\nu}\left(\mathcal{E}_{\mu}^{B} B^{\nu} \right) = 0.
\end{equation}
Note that $\Theta^\mu(\zeta,\lambda,\chi)$ in Eq.~\eqref{total} is the surface term \eqref{eq:surface term} evaluated on the symmetry variations.
From Eq.~\eqref{total}, the off-shell Noether current can be obtained as
\begin{equation}
	\label{eq:off Noether current}
	J_{\rm N}^{\mu}(\zeta,\lambda,\chi) = -\sqrt{-g}\left[ 2\mathcal{E}^{\mu\nu}\zeta_{\nu} + \mathcal{E}_{\nu}^{A}\left( A^{\mu}\zeta^\nu  + \lambda^{\mu\nu} \right) + \mathcal{E}^{B}_{\nu}\left( B^{\mu}\zeta^{\nu} + \chi^{\mu\nu}\right) + \zeta^{\mu}\mathcal{L} \right] + \Theta^{\mu}(\zeta,\lambda,\chi),
\end{equation}
where $J^{\mu}_{\rm N}$ is identically conserved. Thus, the Poincar\'e's lemma guarantees the existence of the off-shell Noether potential $K_{\rm N}^{\mu\nu}$ where $J_{\rm N}^{\mu} = \partial_{\nu}K^{\mu\nu}_{\rm N}$.
Using Eqs.~\eqref{eq:action}, \eqref{eq:eom g}, \eqref{eq:eom aux}, and \eqref{eq:surface term}, we can
obtain the explicit Noether potential:
\begin{equation}
	\label{eq:Noether potential}
	K^{\mu\nu}_{\rm N}(\zeta,\lambda,\chi) = -\sqrt{-g}\left[ 2\alpha\nabla^{[\mu} \zeta^{\nu]} + 4\zeta^{[\mu} \nabla^{\nu]}\alpha + \alpha\left( \left(2 A^{[\mu}\zeta^{\nu]}+\lambda^{\mu\nu} \right) + \beta\left(2 B^{[\mu}\zeta^{\nu]}+\chi^{\mu\nu} \right) \right) \right].
\end{equation}

We are now in a position to obtain quasi-local ADT current by introducing a smooth one-parameter family of solutions $\sigma\in [0,1]$ that interpolates between a reference background at $\sigma = 0$ and the target solution at $\sigma = 1$.
Accordingly, the integration constants in Eqs.~\eqref{eq:metric solution} and \eqref{eq:field solutions} are promoted to $\sigma$-dependent functions,
\begin{equation}
	\label{}
	(\alpha_0, \beta_0, \gamma_0, C, D) \longrightarrow (\alpha_0(\sigma),\beta_0(\sigma),\gamma_0(\sigma), C(\sigma), D(\sigma))
\end{equation}
subject to the boundary conditions:
\begin{align}
	(\alpha_0(0),\beta_0(0),\gamma_0(0), C(0), D(0)) &= (0,0,0,0,0),\label{eq:bdy cond.1}\\
	(\alpha_0(1),\beta_0(1),\gamma_0(1), C(1), D(1)) &= (\alpha_0, \beta_0, \gamma_0, C, D).\label{eq:bdy cond.2}
\end{align}
For the isometry, we take the diffeomorphism generator $\zeta$ to be the timelike Killing vector $\xi = \partial_v$; for the global parts of the gauge symmetries, we choose $\lambda$ and $\chi$ as
\begin{equation}
	\label{eq:global part}
	\lambda^{vr} = -\frac{4}{r^2},\quad  \chi^{vr} = -\frac{1}{4\pi r^2}
\end{equation}
so that $\delta \xi = \delta\lambda=\delta\chi = 0$ along the path in solution space. We also assumed stationarity of the gauge fields,
$\mathcal{L}_{\xi}A^\mu = \mathcal{L}_{\xi}B^\mu = 0$ which gives $\delta(\mathcal{L}_{\xi} g_{\mu\nu}) =\delta(\mathcal{L}_{\xi} A^{\mu})=\delta(\mathcal{L}_{\xi} B^{\mu})=0$ along the path. Thus, it follows that
\begin{equation}
	\label{eq:stationary cond}
	\mathcal{L}_{\xi}\delta g_{\mu\nu} =\mathcal{L}_{\xi}\delta A^{\mu}=\mathcal{L}_{\xi}\delta B^{\mu}=0.
\end{equation}
Varying the off-shell Noether current \eqref{eq:off Noether current} along the path (\textit{i.e.} with respect to $\sigma$) yields
\begin{align}
	\delta J_{\rm N}^{\mu}(\xi,\lambda,\chi) &= -\xi^{\nu}\delta\left[ \sqrt{-g}\left( 2\mathcal{E}^\mu_\nu + \mathcal{E}^{A}_{\nu}A^{\mu} + \mathcal{E}^{B}_{\nu}B^{\mu} \right) \right] \nonumber\\
	&\quad - \partial_{\nu}\left( \delta\left( \alpha\sqrt{-g} \right)\left( \lambda^{\mu\nu} + \chi^{\mu\nu} \right) + \delta\left( \beta\sqrt{-g} \right)\chi^{\mu\nu} \right)\nonumber\\
	&\quad +\sqrt{-g}\left(\mathcal{E}^{\kappa\lambda} \delta g_{\kappa\lambda} - \mathcal{E}_\alpha \delta \alpha - \mathcal{E}_\beta\delta\beta - \mathcal{E}^A_\lambda \delta A^\lambda - \mathcal{E}^B_\lambda\delta B^\lambda\right)\xi^{\mu} \nonumber\\
	&\quad + \delta \Theta^\mu(\xi,0,0)- \mathcal{L}_{\xi}\Theta^{\mu}(\delta g, \delta A, \delta B) - 2\partial_{\nu}\left( \xi^{[\mu}\Theta^{\nu]}(\delta g, \delta A, \delta B) \right).\label{eq:var Noether J}
\end{align}
Using Eq.~\eqref{eq:stationary cond}, the first two terms in fourth line in Eq.~\eqref{eq:var Noether J} vanish: $\delta \Theta^\mu(\xi,0,0) = \mathcal{L}_{\xi}\Theta^{\mu}(\delta g, \delta A, \delta B) = 0$.
Let us now consider the off-shell ADT current defined in Ref.~\cite{Kim:2013zha}
\begin{equation}
\label{offshellADTcurrent}
J_{\rm ADT}^{\mu} = \delta J_{\rm N}^{\mu}(\xi,\lambda,\chi) + 2\partial_{\nu}\left( \xi^{[\mu}\Theta^{\nu]} (\delta g, \delta A, \delta B)\right)
\end{equation}
where $\partial_{\mu}J_{\rm ADT}^{\mu} =0$.
From Eq.~\eqref{eq:var Noether J}, we obtain the ADT current as
\begin{align}
	\label{offshellADT}
	J^{\mu}_{\rm ADT}(\xi,\lambda,\chi) &= -\xi^{\nu}\delta\left[ \sqrt{-g}\left( 2\mathcal{E}^\mu_\nu + \mathcal{E}^{A}_{\nu}A^{\mu} + \mathcal{E}^{B}_{\nu}B^{\mu} \right) \right] \nonumber\\
	&\quad - \partial_{\nu}\left( \delta\left( \alpha\sqrt{-g} \right)\left( \lambda^{\mu\nu} + \chi^{\mu\nu} \right) + \delta\left( \beta\sqrt{-g} \right)\chi^{\mu\nu} \right)\nonumber\\
	&\quad +\sqrt{-g}\left(\mathcal{E}^{\kappa\lambda} \delta g_{\kappa\lambda} - \mathcal{E}_\alpha \delta \alpha - \mathcal{E}_\beta\delta\beta - \mathcal{E}^A_\lambda \delta A^\lambda - \mathcal{E}^B_\lambda\delta B^\lambda\right)\xi^{\mu}.
\end{align}

Next, we consider the ADT potential defined through $J_{\rm ADT}^{\mu} = \partial_{\nu}K^{\mu\nu}_{\rm ADT}$, then the quasi-local ADT charge associated with $\{\xi, \lambda, \chi\}$, linearized along the parameter $\sigma$, is obtained as
\begin{align}
	\label{eq:dQ}
	\delta Q[\xi, \lambda, \chi] &= \int_{\Sigma} \dd[2]{x}_{\mu\nu} K^{\mu\nu}_{\rm ADT}(\xi, \lambda, \chi) \nonumber\\
	&= \int_{\Sigma} \dd[2]{x}_{\mu\nu}\left[ \delta K^{\mu\nu}_{\rm N}(\xi, \lambda, \chi) +  2\xi^{[\mu}\Theta^{\nu]}(\delta g, \delta A, \delta B) \right],
\end{align}
where $\delta K^{\mu\nu}_{\rm N}(\xi, \lambda, \chi)$ is the variation of Eq.~\eqref{eq:Noether potential} with respect to $\sigma$ for $\zeta = \xi$. Here, $\dd[2]x_{\mu\nu} = \frac{1}{2}\epsilon_{\mu\nu\kappa\lambda}\dd x^\kappa \wedge \dd x^\lambda$ along with $\epsilon_{vr\theta\phi} = 1$, and  $\Sigma$ is a two-sphere at fixed radius $r$ outside the horizon.
Explicitly, substituting Eqs.~\eqref{eq:surface term}, \eqref{eq:metric solution}, \eqref{eq:field solutions} into Eq.~\eqref{eq:dQ} yields
\begin{align}
	\delta Q[\xi, 0, 0] &= 4\pi\left[ \left( C(\sigma) + \gamma_0(\sigma) \right)\partial_{\sigma}\alpha_0(\sigma) + D(\sigma)\partial_{\sigma}\left( \alpha_0(\sigma)\beta_0(\sigma) \right)+2\alpha_0(\sigma)\partial_{\sigma}\gamma_{0}(\sigma) \right],\label{eq:Q xi}\\
	\delta Q[0, \lambda, 0] &= 16\pi \partial_{\sigma} \alpha_{0}(\sigma),\label{eq:Q lambda}\\
	\delta Q[0, 0, \chi] &= \partial_{\sigma}\left( \alpha_0(\sigma)\beta_{0}(\sigma) \right).\label{eq:Q chi}
\end{align}
Equation~\eqref{eq:Q xi} is the pullback to the interpolation path of the following one-form on the parameter space of solutions:
\begin{equation*}
	\delta Q[\xi,0,0]
	=4\pi\left[(C+\gamma_0)\delta\alpha_0
	+D\delta(\alpha_0\beta_0)
	+2\alpha_0\delta\gamma_0\right].
\end{equation*}
Its exterior derivative is
\begin{equation}
	\label{eq:charge-integrability}
	\delta \left(\delta Q[\xi,0,0]\right)
	=4\pi\left[\delta(C-\gamma_0)\wedge\delta\alpha_0
	+\delta D\wedge\delta(\alpha_0\beta_0)\right].
\end{equation}

A simple and natural way to ensure integrability while retaining arbitrary independent variations of $\alpha_0$, $\beta_0$, and $\gamma_0$ is to restrict the solution phase space to the sector
\begin{equation}
	\label{eq:integrable-sector}
	\delta(C-\gamma_0)=0,\qquad \delta D=0.
\end{equation}
Since the reference solution in Eq.~\eqref{eq:bdy cond.1} satisfies $C(0)=\gamma_0(0)=D(0)=0$, the corresponding fixed values vanish, and hence
\begin{equation}
	\label{eq:integrable-path}
	C(\sigma)=\gamma_0(\sigma),\qquad D(\sigma)=0.
\end{equation}
These conditions select an integrable sector of the solution phase space rather than imposing a boundary condition at a particular spacetime surface. In particular, $C$ and $D$ are homogeneous integration constants of the auxiliary gauge fields that are not fixed by the local equations of motion alone. Along the interpolation path, Eq.~\eqref{eq:Q xi} then reduces to
\begin{equation}
	\label{eq:exact-mass-charge}
	\delta Q[\xi,0,0]=8\pi\partial_{\sigma}\left(\alpha_0(\sigma)\gamma_0(\sigma)\right),
\end{equation}
which is the pullback of the exact one-form $8\pi\delta(\alpha_0\gamma_0)$ and is therefore independent of the interpolation path.

Integrating Eqs.~\eqref{eq:Q xi},~\eqref{eq:Q lambda}, and \eqref{eq:Q chi} along $\sigma\in [0,1]$ with the boundary conditions \eqref{eq:bdy cond.1} and \eqref{eq:bdy cond.2} gives
\begin{align}
	Q[\xi,0,0] &= \int^1_0  \delta Q\left[ \xi,0,0 \right] \dd \sigma= 8\pi\int^1_0\partial_{\sigma}\left( \alpha_0(\sigma) \gamma_0(\sigma) \right) \dd \sigma = 8\pi\alpha_0 \gamma_0, \label{q1}\\
	Q[0,\lambda,0] &= \int^1_0  \delta Q\left[0,\lambda,0 \right] \dd \sigma= 16\pi\alpha_0,\\
	Q[0,0,\chi] &= \int^1_0 \delta Q\left[0,0,\chi \right] \dd \sigma= \alpha_0\beta_0.
\end{align}
Finally, if we identify $\alpha_0 = \frac{1}{16\pi G}$, $\beta_0 =2\Lambda$, and $\gamma_0 = 2G M$, the three conserved charges can be neatly written as
\begin{equation}
\label{eq:conserved quant}
	Q[\xi,0,0] = M,\quad Q[0,\lambda,0] = \frac{1}{G},\quad Q[0,0,\chi] = \frac{\Lambda}{8\pi G}.
\end{equation}
In these identifications, the solutions \eqref{eq:metric solution} and \eqref{eq:field solutions} are also written as
\begin{gather}
	\label{metricsolution}
		\dd s^2 = -f(r)\dd v^2 + 2\dd v\dd r + r^2 \dd \Omega^2, \quad f(r)= 1-\frac{2G M}{r} + \frac{\Lambda}{3}r^{2}, \\
	A^r = -\frac{4}{3} \Lambda r+ \frac{2G M}{r^2}, \quad B^r = \frac{1}{3}r. \label{vectorpotential}
\end{gather}
Thus, it turns out that the gravitational constant can be realized as a quasi-local conserved charge associated with the gauge symmetry of $A^{\mu}$.

Note that the conserved quantity $G$ is assumed to be a positive, non-zero constant.
From $Q[0,\lambda,0] = G^{-1}$ in Eq.~\eqref{eq:conserved quant}, the limit $G\to 0$ makes the charge diverge and hence ill-defined.
Moreover, to ensure attractive gravity, we require $G>0$.
More precisely, in the limit $\Lambda \to 0$ the solution reduces to the Schwarzschild geometry which is fixed by setting $\gamma_0 = 2GM >0$.
If $G<0$, this would instead force $M<0$, which is unphysical.
Therefore, we keep $G$ to be a positive, non-zero constant.

As a comment, the auxiliary fields serve as Lagrange multiplier type variables so that they are non-propagating.
The appearance of $G$ and $\Lambda$ as conserved quantities is a structural consequence of the enlarged action, rather than an indication of new gravitational dynamics.	

\section{Extended thermodynamic first law and Smarr formula}
\label{sec:extended}
In the off-shell ADT current \eqref{offshellADT} conserved identically, let us consider a four-volume $\mathcal{V}$ bounded by two Cauchy hypersurfaces $\Sigma_{v}$ and $\Sigma_{v'}$, and a timelike boundary $\mathcal{T}$, then the Stokes' theorem
tells us that
\begin{equation}
	\label{}
	\int_{\mathcal{V}} \dd[4]{x} \partial_{\mu}J^{\mu}_{\rm ADT} = \oint_{\partial \mathcal{V}} \dd[3]{x_\mu} J^{\mu}_{\rm ADT}= \int_{\Sigma_v} \dd[3]{x_{\mu}} J^{\mu}_{\rm ADT} - \int_{\Sigma_{v'}} \dd[3]{x_{\mu}} J^{\mu}_{\rm ADT} + \int_{\mathcal{T}} \dd[3]{x_{\mu}} J^{\mu}_{\rm ADT} = 0,
\end{equation}
where $\dd[3]x_{\mu} = \frac{1}{3!}\epsilon_{\mu\alpha\beta\gamma}\dd x^\alpha \wedge \dd x^\beta \wedge \dd x^\gamma$.
Next, assuming there does not exist ADT flux through the timelike boundary, \textit{i.e.},  $\int_{\mathcal{T}} \dd[3]{x_{\mu}} J^{\mu}_{\rm ADT} = 0$,
one can obtain $\int_{\Sigma_v} \dd[3]{x_{\mu}} J^{\mu}_{\rm ADT}=\int_{\Sigma_{v'}} \dd[3]{x_{\mu}} J^{\mu}_{\rm ADT}$.
Let $\Sigma_{v}$ be a (partial) Cauchy hypersurface whose boundary $\partial \Sigma_{v}$ consists of two 2-sphere cross sections at advanced time $v$: one is at finite radius $r$ outside the horizon (denoted by $\Sigma$) and the other is on the Killing horizon (denoted by $\mathcal{H}$).
Additionally, assuming there does not exist ADT sources in the interior of $\Sigma_{v}$, one can obtain
\begin{equation}
	\label{ADTzero}
	\int_{\Sigma_v} \dd[3]{x_{\mu}} J^{\mu}_{\rm ADT} = 0
\end{equation}
for any $v$.
Using $J_{\rm ADT}^{\mu} = \partial_{\nu}K^{\mu\nu}_{\rm ADT}$ and Stokes' theorem on $\Sigma_v$, we find that Eq.~\eqref{ADTzero} becomes
\begin{equation}
	\label{}
	\int_{\Sigma_v} \dd[3]{x_{\mu}} \partial_{\nu} K^{\mu\nu}_{\rm ADT} = \oint_{\partial\Sigma_v} \dd[2]{x_{\mu\nu}}  K^{\mu\nu}_{\rm ADT} = \int_{\Sigma} \dd[2]{x_{\mu\nu}} K^{\mu\nu}_{\rm ADT} - \int_{\mathcal{H}} \dd[2]{x_{\mu\nu}} K^{\mu\nu}_{\rm ADT} = 0,
\end{equation}
which implies
\begin{equation}
	\label{equal}
	\int_{\Sigma} \dd[2]{x_{\mu\nu}} K^{\mu\nu}_{\rm ADT} = \int_{\mathcal{H}} \dd[2]{x_{\mu\nu}} K^{\mu\nu}_{\rm ADT}.
\end{equation}
By using Eq.~\eqref{equal}, the infinitesimal variation~\eqref{eq:dQ} at finite radius for the isometry can be computed  in terms of horizon integrals:
\begin{align}
		\delta Q[\xi,0,0] &= \int_{\Sigma} \dd[2]{x}_{\mu\nu} K^{\mu\nu}_{\rm ADT}(\xi, 0,0) \nonumber\\
		&= \int_{\mathcal{H}} \dd[2]{x}_{\mu\nu}   \delta K^{\mu\nu}_{\rm N}(\xi, 0,0) + 2\int_{\mathcal{H}} \dd[2]{x}_{\mu\nu}\xi^{[\mu}\Theta^{\nu]} (\delta g, \delta A, \delta B).\label{eq:ADT at horizon}
\end{align}

To describe the motion of the horizon under variations, we introduce $\eta = (\delta r_h) \partial_{r}$, where $r_h$ is the horizon radius.
The deformation vector $\eta$ shifts the location of the horizon.
By the Reynolds transport theorem~\cite{Frankel:1997ec}, one can find
$\delta\left( \int_{\mathcal{H}} \dd[2]{x}_{\mu\nu} K^{\mu\nu}_{\rm N}(\xi, 0,0) \right) = \int_{\mathcal{H}}   \dd[2]{x}_{\mu\nu} \delta K^{\mu\nu}_{\rm N}(\xi, 0,0) + \int_{\mathcal{H}}\mathcal{L}_{\eta}\left( \dd[2]{x}_{\mu\nu} K^{\mu\nu}_{\rm N}(\xi, 0,0) \right)$ so that the first term in Eq.~\eqref{eq:ADT at horizon}
can be written in terms of two horizon integrals:
\begin{equation}
	\label{eq:var delta K}
	\int_{\mathcal{H}} \dd[2]{x}_{\mu\nu}  \delta K^{\mu\nu}_{\rm N}(\xi, 0,0)  = \delta\left( \int_{\mathcal{H}} \dd[2]{x}_{\mu\nu} K^{\mu\nu}_{\rm N}(\xi, 0,0) \right) - \int_{\mathcal{H}} \mathcal{L}_{\eta}\left( \dd[2]{x}_{\mu\nu} K^{\mu\nu}_{\rm N}(\xi, 0,0) \right).
\end{equation}
To compute the first term on the right-hand side of Eq.~\eqref{eq:var delta K},
we rewrite Eq.~\eqref{eq:Noether potential} for the isometry on the horizon as
\begin{align}
	\label{eq:Noether potential integral}
	\int_{\mathcal{H}} \dd[2]{x}_{\mu\nu} K^{\mu\nu}_{\rm N}(\xi, 0,0) &= -\int_{\mathcal{H}} \dd \mathcal{A} (\xi_{\mu}n_{\nu}- \xi_{\nu}n_{\mu}) \left( \alpha\nabla^{[\mu} \xi^{\nu]} + 2\xi^{[\mu} \nabla^{\nu]}\alpha + \alpha \left( A^{[\mu}\xi^{\nu]}  + \beta B^{[\mu}\xi^{\nu]} \right) \right)\nonumber\\
	&= \int_{\mathcal{H}}  \left( 2\kappa \alpha -2\xi^{\mu}\nabla_{\mu}\alpha+ \alpha \left( \xi_{\mu}A^\mu + \beta\xi_{\mu}B^\mu \right) \right)\dd\mathcal{A}.
\end{align}
Here, $\kappa = \eval{-n_\mu \xi^\nu\nabla_{\nu}\xi^\mu}_{\mathcal{H}}$ is the surface gravity, $n=n^\mu \partial_{\mu} = -\partial_r$ is a future-directed null vector field satisfying $n^\mu n_\mu = 0$ and $\xi^\mu n_\mu = -1$,
and
$\dd[2]{x_{\mu\nu}} = \frac{1}{2}\sqrt{-g}\epsilon_{\mu\nu\kappa\lambda}\dd x^\kappa \wedge \dd x^\lambda = \frac{1}{2}(\xi_{\mu}n_{\nu}- \xi_{\nu}n_{\mu})\dd \mathcal{A}$ with $\dd \mathcal{A} = r^{2}\sin\theta \dd \theta \dd \phi$.
Plugging the solutions \eqref{eq:metric solution} and \eqref{eq:field solutions} into Eq.~\eqref{eq:Noether potential integral}, we obtain
\begin{equation}
	\label{ADTpotential at the horizon}
	\int_{\mathcal{H}} \dd[2]{x}_{\mu\nu} K^{\mu\nu}_{\rm N}(\xi, 0,0) = TS + 16\pi\alpha_0 \Phi + \alpha_0\beta_0 V,
\end{equation}
where we used the definitions:
\begin{equation}
	\label{eq:therm vari def}
	T = \frac{\kappa}{2\pi}, \quad S = 4\pi \alpha_0  \int_{\mathcal{H}}\dd\mathcal{A}, \quad \Phi = \frac{1}{16\pi}\int_{\mathcal{H}} \xi_{\mu}A^\mu\dd\mathcal{A}, \quad  V = \int_{\mathcal{H}} \xi_{\mu}B^\mu \dd\mathcal{A}.
\end{equation}
For Eqs.~\eqref{eq:metric solution} and \eqref{eq:field solutions}, the second term on the right-hand side of Eq.~\eqref{eq:var delta K} is easily calculated as
\begin{equation}
	\label{eq:horizon shift}
	\int_{\mathcal{H}} \mathcal{L}_{\eta}\left( \dd[2]{x}_{\mu\nu} K^{\mu\nu}_{\rm N}(\xi, 0,0) \right)=  \int_{\mathcal{H}} \dd[2]{x}_{\mu\nu} \left( \eta^{\lambda}\nabla_{\lambda}K^{\mu\nu}_{\rm N} - \eta^\lambda \Gamma_{\rho\lambda}^\rho K^{\mu\nu}_{\rm N} \right) = 0.
\end{equation}
Combining Eqs.~\eqref{ADTpotential at the horizon} and \eqref{eq:horizon shift}, we can rewrite the first term in Eq.~\eqref{eq:ADT at horizon} as
\begin{equation}
	\label{eq:Noether var}
	\int_{\mathcal{H}} \dd[2]{x}_{\mu\nu} \delta K^{\mu\nu}_{\rm N}(\xi, 0,0) = \delta \left( TS + 16\pi\alpha_0\Phi + \alpha_0\beta_0 V\right). 
\end{equation}
The second horizon term in Eq.~\eqref{eq:ADT at horizon} is also computed as~\cite{Compere:2006my}
\begin{equation}
	\label{eq:remain var}
	2\int_{\mathcal{H}}\dd[2]{x}_{\mu\nu} \xi^{[\mu}\Theta^{\nu]}(\delta g, \delta A, \delta B) = -S\delta T - 16\pi\alpha_0 \delta \Phi - \alpha_0\beta_0 \delta V.
\end{equation}
Plugging Eqs.~\eqref{eq:Noether var} and \eqref{eq:remain var} into Eq.~\eqref{eq:ADT at horizon}, we obtain
\begin{equation}
	\label{eq:1st law prototype}
	\delta Q[\xi,0,0] = T\delta S + \Phi \delta\left( 16\pi \alpha_0 \right) + V \delta \left( \alpha_0\beta_0 \right).
\end{equation}
For the solutions \eqref{metricsolution} and \eqref{vectorpotential}, the variables in Eq.~\eqref{eq:therm vari def} can be expressed by
\begin{equation}
	\label{}
	T = \frac{1}{4\pi}\left( \frac{1}{r_h}+\Lambda r_h \right),\quad S = \frac{\pi r_h^2}{G},\quad \Phi = \frac{1}{4}(r_h- \Lambda r_h^3),\quad V = \frac{4}{3}\pi r_h^3,
\end{equation}
along with $M = \frac{1}{2G}\left( r_h + \frac{\Lambda}{3}r_h^3 \right) $ and $P = - \frac{\Lambda}{8\pi G}$.
Plugging these variables into Eq.~\eqref{eq:1st law prototype}, we obtain the extended thermodynamic first law:
\begin{equation}
	\label{eq:1st law}
	\delta M = T\delta S + \Phi \delta G^{-1} - V \delta P.
\end{equation}
Therefore, it turns out that the gravitational constant can play a role of a thermodynamic variable in the extended first law.

Finally, we derive the Smarr formula
based on the following scaling of thermodynamic variables $(S, G^{-1}, P)$ ~\cite{Smarr:1972kt}
\begin{equation}
	\label{}
	S \longrightarrow a S,\quad G^{-1} \longrightarrow a G^{-1},\quad P \longrightarrow a P,
\end{equation}
where $a$ is a dimensionless parameter.
Under this scaling, the mass scales linearly as
\begin{equation}
	\label{}
	M\left( S, G^{-1}, P \right) \longrightarrow M\left( a S, a G^{-1}, aP \right) = a M\left( S, G^{-1}, P \right),
\end{equation}
and so $M$ is homogeneous of degree one.
Euler's theorem then gives
\begin{equation}
	\label{Smarr}
	M = \left( \pdv{M}{S} \right)_{G^{-1},P}S+\left( \pdv{M}{G^{-1}} \right)_{S,P}G^{-1} + \left( \pdv{M}{P} \right)_{S,G^{-1}}P,
\end{equation}
where $\left( \pdv{M}{S} \right)_{G^{-1},P} = T,\quad \left( \pdv{M}{G^{-1}} \right)_{S,P} =  \Phi,\quad \left( \pdv{M}{P} \right)_{S,G^{-1}} = -V$
from the thermodynamic first law \eqref{eq:1st law}.
Thus, the Smarr formula can be obtained as
\begin{equation}
	\label{eq:Smarr}
	M = TS + \Phi G^{-1} - PV,
\end{equation}
which is compatible with Eq.~\eqref{ADTpotential at the horizon} using the identifications of $G^{-1} = 16\pi \alpha_0$ and $P = -\alpha_0 \beta_0$.
\section{Conclusion and discussion}
\label{sec:conclusion}
In a covariant off-shell quasi-local ADT framework, we have shown that the gravitational constant can be realized as a conserved charge generated by the global component of a gauge symmetry via a scalar–gauge pair.
This construction identifies the gravitational constant as a conserved quantity associated with the gauge symmetry of the auxiliary fields.
In addition, we found the extended thermodynamic first law and the Smarr formula.

The current work naturally raises a question: ``Where is the gravitational constant as a charge in physical space?'' For simplicity, if we consider the starting action \eqref{eq:action} without the cosmological constant by taking $\beta \to 0$, the vacuum solutions reduce to the metric function $f(r)=1-\frac{2GM}{r}$ in Eq.~\eqref{metricsolution} and the vector potential $A^r=\frac{2GM}{r^2}$ in Eq.~\eqref{vectorpotential}. In Eq.~\eqref{eq:eom aux}, the equation of motion relating the vector potential to the Ricci scalar $R$ is $\nabla_{\mu}A^\mu = R$. For $r\neq 0$, the physical field tensor $F=\nabla_{\mu}A^\mu$ and the Ricci scalar vanish.
To investigate the behaviour around $r=0$, we use the Noether potential \eqref{eq:Noether potential} together with $\lambda^{\mu\nu}$ in Eq.~\eqref{eq:global part}.
Then, the Noether charge can be evaluated as $\int_{\Sigma} K_{\rm N}^{\mu\nu}(0,\lambda,0)\,\dd[2]{x}_{\mu\nu} = -\int_{\Sigma} \alpha\,\lambda^{\mu\nu} \sqrt{-g}\,\dd[2]{x}_{\mu\nu} = G^{-1}$, with $\Sigma$ being a 2-sphere at
an arbitrary finite radius outside the horizon. This implies that the gravitational constant
as a source, along with the mass, is concentrated at $r=0$ like the electrostatic point charge located at the origin.

One might wonder whether the gravitational constant can be interpreted as hair of a black hole. Although $G^{-1}$ appears as a thermodynamic charge in the extended first law, its physical status differs from that of the mass. The common first-law structure means that $M$ and $G^{-1}$ are both coordinates on the extended space of stationary solutions; it does not imply that they have the same dynamical origin or operational status.

The mass $M=Q[\xi,0,0]$ is the Hamiltonian charge associated with the timelike diffeomorphism generated by $\xi$. It labels individual black-hole states in a fixed theory and can change through physical energy fluxes. In contrast, $G^{-1}=Q[0,\lambda,0]$ is associated with the global part of a gauge symmetry of the auxiliary field. The auxiliary equation of motion $\nabla_\mu\alpha=0$ constrains $\alpha=(16\pi G)^{-1}$ to be constant, and the auxiliary sector contains no local propagating degree of freedom that could carry this charge.

Accordingly, a variation of $G^{-1}$ in the extended first law represents a comparison between neighbouring solutions or coupling sectors in the extended phase space, rather than a radiative evolution of a black hole within a theory with fixed $G$. For an ordinary physical process in a fixed gravitational theory, one restricts the variation to $\delta G^{-1}=0$. Thus, $G^{-1}$ is a genuine conserved charge of the enlarged theory and may be regarded as a thermodynamic variable, but it is not standard black-hole hair: it is universal within a fixed coupling sector and does not define an independent radiative attribute of an individual black hole. In particular, there is no propagating auxiliary mode through which this charge could be carried away, including through Hawking radiation.

\acknowledgments
We would like to thank Sojeong Cheong for exciting discussions. This research was supported by Basic Science Research Program through the National Research Foundation of Korea (NRF) funded by the Ministry of Education through the Center for Quantum Spacetime (CQUeST) of Sogang University (No. RS-2020-NR049598).
This work was supported by the National Research Foundation of Korea (NRF) grant funded by the Korea government (MSIT) (No. RS-2022-NR069013).


\bibliographystyle{JHEP}       

\bibliography{reference}

\end{document}